\documentclass[aps,prl,floatfix,twocolumn,superscriptaddress]{revtex4}
\usepackage{latexsym}
\usepackage{graphicx}
\usepackage{rotating}
\usepackage{hyperref}
\usepackage{amsmath,amssymb,amsfonts}
\usepackage{bm}
\usepackage{color}

\newcommand{\emphasize}{\emph}

\def\onlinecite#1{\cite{#1}}

\newcommand{\norm}[1]{\ensuremath{| #1 |}}

\newcommand{\ba}{BaFe$_2$As$_2$ }
\newcommand{\se}{KFe$_2$Se$_2$ }

\begin{document}

%\title{Order-by-disorder selection quenched by magnetic impurities unveils novel magnetic order in pnictides}
%\title{Competition between collinear and 90 degree order induced by impurities in the frustrated Heisenberg model of pnictides}
%\title{Quenching of order-by-disorder by non-magnetic impurities in the frustrated Heisenberg model of pnictides}
%\title{Non-collinear magnetic phase induced by impurity-quenching of order-by-disorder in the frustrated Heisenberg model of pnictides}
   \title{Anticollinear magnetic order induced by impurities in the frustrated Heisenberg model of pnictides}
%\title{Quenching of order-by-disorder with impurities: A novel magnetic phase in the frustrated Heisenberg model of pnictides}
%\title{Non-collinear magnetic order induced by impurities in the frustrated Heisenberg model of pnictides}
%\title{90 degree magnetic order induced by impurities in the frustrated Heisenberg model of pnictides}
%\title{Beating order-by-disorder with impurities: 90 degree magnetic order in the frustrated Heisenberg model of pnictides.}
%\title{Competition between thermal order-by-disorder and 90 degree magnetic order induced by impurities in the frustrated Heisenberg model of pnictides}
%\title{Quenching of order-by-disorder and stabilization of a novel magnetic order in the frustrated Heisneberg model of pnictides.}

\author{C\'edric Weber$^{*}$}
\affiliation{Cavendish Laboratory, J.J. Thomson Ave, Cambridge CB3 0HE, U.K.}
\author{Fr\'ed\'eric Mila}
\affiliation{Institute of Theoretical Physics,
\'{E}cole Polytechnique F\'{e}d\'{e}rale de Lausanne (EPFL),
Station 12, 1015 Lausanne, Switzerland}

%%%%%%%%%%%%%%%%%%%%%%%%%%%%%%%%%%%%%%%%%%%%%%%%%%%%%%%%%%%%%%%%
%%%%%%%%%%%%%%%%%%%%%%%%%%%%%%%%%%%%%%%%%%%%%%%%%%%%%%%%%%%%%%%%

\begin{abstract}
We present Monte Carlo simulations for a classical antiferromagnetic
Heisenberg model with both nearest ($J_1$) and next-nearest ($J_2$) exchange
couplings on the square lattice in the presence of non-magnetic impurities.
We show that the order-by-disorder entropy selection, associated with the Ising-like phase transition that appears for $J_2/J_1>1/2$ in the
pure spin model, is quenched at low temperature due to the presence of non-magnetic impurities.
Evidences that a new competing order is stabilized around the impurities, and in turn induces a re-entrance phase
transition are reported. Implications for local magnetic measurement of the parent compound of iron pnictides are briefly discussed.
\end{abstract}

\maketitle

%%%%%%%%%%%%%%%%%%%%%%%%%%%%%%%%%%%%%%%%%%%%%%%%%%%%%%%%%%%%%%%%
%%%%%%%%%%%%%%%%%%%%%%%%%%%%%%%%%%%%%%%%%%%%%%%%%%%%%%%%%%%%%%%%
%%%%%%%%%%%%%%%%%%%%%%%%%%%%%%%%%%%%%%%%%%%%%%%%%%%%%%%%%%%%%%%%
%%%%%%%%%%%%%%%%%%%%%%%%%%%%%%%%%%%%%%%%%%%%%%%%%%%%%%%%%%%%%%%%
%%%%%%%%%%%%%%%%%%%%%%%%%%%%%%%%%%%%%%%%%%%%%%%%%%%%%%%%%%%%%%%%
%%%%%%%%%%%%%%%%%%%%%%%%%%%%%%%%%%%%%%%%%%%%%%%%%%%%%%%%%%%%%%%%
%%%%%%%%%%%%%%%%%%%%%%%%%%%%%%%%%%%%%%%%%%%%%%%%%%%%%%%%%%%%%%%%
%%%%%%%%%%%%%%%%%%%%%%%%%%%%%%%%%%%%%%%%%%%%%%%%%%%%%%%%%%%%%%%%

Unconventional superconductivity occurs in the proximity of magnetically ordered state in many materials \cite{spin_disorder_ref1,spin_disorder_ref2}.
Understanding the magnetic phase of the parent compound is an important step towards understanding the mechanism of superconductivity.
Unlike the case of the cuprates, magnetism and its underlying electronic state in the iron pnictide superconductor \ba \cite{spin_disorder_ref3} is still not well understood.
Many low-energy probes such as transport \cite{spin_disorder_ref_resistivity}, scanning tunnelling microscopy \cite{spin_disorder_ref_nemacity} and angle-resolved photoemission spectroscopy \cite{spin_disorder_ref_dirac_cone} have measured strong in-plane anisotropy of
the electronic states, but there is no consensus on its physical origin.

It was suggested from first principle calculations \cite{pnictides_orbital_order_and_no_magnetic} that the origin stems from orbital order,
although the obtained anisotropy in the resistivity is opposite to the one found
experimentally \cite{pnictides_wrong_anisotropy_orbital_order}.
A more likely scenario is related to spin density wave instabilities, which is supported
by recent neutron diffraction measurements \cite{spin_disorder_incommensurate_magnetism}, and stems from Fermi surface nesting of electron and hole pockets.
In the latter picture, the nematic magnetic order introduce in turn an orbital polarization, since the electron pockets at $\bold{Q}=(\pi,0)$ and $\bold{Q}=(0,\pi)$ have
$d_{yz}$ and $d_{xz}$ orbital characters respectively \cite{pnictides_anisotropy_is_magnetic_2,pnictides_anisotropy_is_magnetic}.
The exact nature of the magnetic ground state remains however unclear.
It was suggested that the commensurate AFM order can also be described within a local moment picture
that may become relevant in the presence of moderately large electronic correlations and can be quantified, for example, in terms of the Heisenberg model
with both nearest- (J$_1$) and next-nearest (J$_2$) exchange couplings in the frustrated regime $J_2 > 2J_1$ \cite{spin_disorder_j2j1}.
The first estimation of the coupling constants J$_{1,2}$, obtained by fitting the experimental spin density wave excitation spectra,
yielded parameters which are not in the frustrated regime~\cite{pnictides_spin_wave_fit_too_small_J}.
However, it was later shown that the fits of the experimental data included
energy scales beyond 100meV, which are not well described by magnon excitations~\cite{pnictides_argument_fit_of_nature_is_large_energy}.
A more careful study, including the itinerant character of the electrons,
suggested that indeed the pnictides are in the frustrated regime \cite{pnictides_ilya_frustration_is_large}.
This scenario was also recently also supported by first-principle calculations for selenium based compounds (\se) \cite{spin_disorder_electronic_structure},
and the iron pnictides \ba and \se compounds were proposed as experimental realisations of a layered $J_1{-}J_2$ spin model in the collinear regime \cite{chandra_collin}.
In particular, it has been suggested both experimentally \cite{spin_disorder_nmr_impurities} and theoretically \cite{spin_disorder_ref_theory_impurities} that non-magnetic
impurities have a dramatic impact on the magnetic and superconducting properties.

In this Letter, we address the question of the interplay between the frustration, induced by the exchange coupling, and the disorder induced by the imperfections of the crystallographic
structure. Firstly, we consider non-magnetic impurities, and then extend the calculations to magnetic impurities.
Since density functional calculations, and quite generally quantum based calculations,
are limited to relatively small unit-cell and cannot tackle the issue of large super-cell structures,
we limit our calculations to the classical case, and carry out Monte Carlo calculations of the $J_1{-}J_2$ model in the presence
of non-magnetic impurities. The methodology and implementation were discussed in Ref.~\onlinecite{cedric,cedric_phonon}, a short summary is given hereafter.

The Heisenberg hamiltonian reads:
\begin{equation}\label{ham}
\hat{\cal{H}} = \sum_{\langle i,j \rangle} J_1
\hat{{\bf {S}}}_{i} \cdot \hat{{\bf {S}}}_{j}
+ \sum_{\langle \langle i,j \rangle \rangle} J_2
\hat{{\bf {S}}}_{i} \cdot \hat{{\bf {S}}}_{j},
\end{equation}
where $\hat{{\bf {S}}}_{i}$ are O(3) spins on a periodic square lattice
with $N=L \times L$ sites. $\langle i,j \rangle$ and
$\langle \langle i,j \rangle \rangle$ indicate the sum over nearest and
next-nearest neighbors, respectively. $J_1$ sets the energy scale, and
$J_2/J_1=0.55$ is used throughout the rest of the paper.
In absence of disorder, the magnetic vector is ${\bf Q}=(\pi,\pi)$
for $J_2/J_1<0.5$, and  for $J_2/J_1>0.5$
the ground state is continuously degenerate, but the
entropy selection reduces the O(3) symmetry of the ground state to Z$_2$
at finite temperature, selecting the states with antiferromagnetic spin correlations in
one spatial direction and ferromagnetic correlations in the other (${\bf Q}=(0,\pi),(\pi,0)$).
This is the so-called \emphasize{order by disorder} scenario, and the associated discrete symmetry
breaking drives a finite temperature Ising-like phase transition \cite{chandra_collin}.
To characterize this transition,
it is useful to construct, from the original spin variables
$\hat{{\bf {S}}}_{i}$, an effective Ising variable on the dual lattice:
\begin{equation}
\label{isingvar}
M_2(x)=
(\hat{{\bf {S}}}_{i} - \hat{{\bf {S}}}_{k}) \cdot
(\hat{{\bf {S}}}_{j} - \hat{{\bf {S}}}_{l}),
\end{equation}
where $(i,j,k,l)$ are the corners with diagonal $(i,k)$ and $(j,l)$ of the
plaquette centered at the site $x$ of the dual lattice, and we define
its normalized counterpart as $Z_2(x)=M_2(x)/|M_2(x)|$.
In this way, the two collinear states with ${\bf Q}=(\pi,0)$ and
${\bf Q}=(0,\pi)$ can be distinguished by the value of the
Ising variable, $Z_2(x)= \pm 1$.

%%%%%%%%%%%%%%%%%%%%%%%%%%%%%%%%%%%%%%%%%%%%%%%%%%%%%%%%%%%%%%%%
%%%%%%%%%%%%%%%%%%%%%%%%%%%%%%%%%%%%%%%%%%%%%%%%%%%%%%%%%%%%%%%%
%%%%%%%%%%%%%%%%%%%%%%%%%%%%%%%%%%%%%%%%%%%%%%%%%%%%%%%%%%%%%%%%
%%%%%%%%%%%%%%%%%%%%%%%%%%%%%%%%%%%%%%%%%%%%%%%%%%%%%%%%%%%%%%%%

\begin{figure}
\begin{center}
\includegraphics[width=1.0\columnwidth]{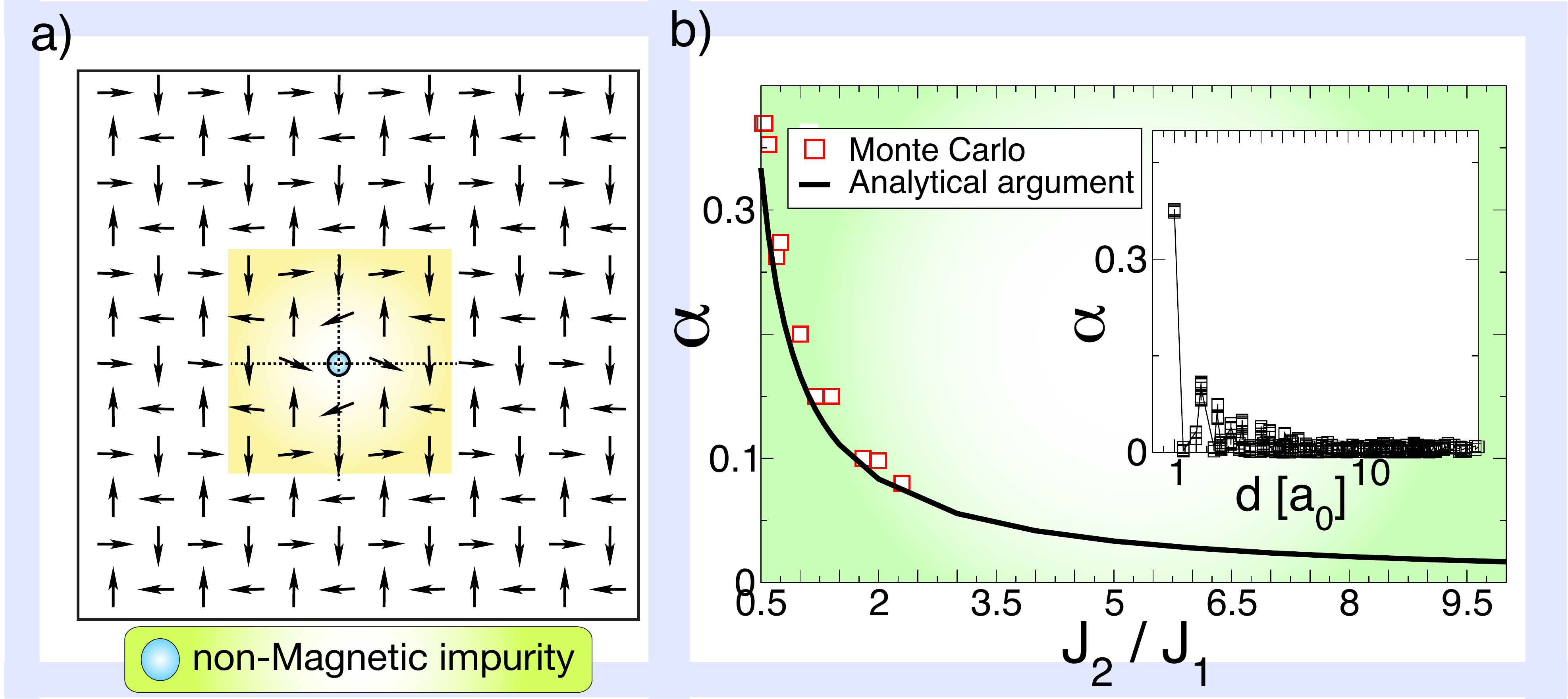}
\caption{
(Color online) a) Typical spin configuration obtained at $T=10^{-6}J_1$. The filled circle indicates the location
of the impurity, and the rectangle highlights the region around the impurity where the spins deviate from the $90^\circ$ ordered state for $J_2/J_1=0.55$.
b) Angle between the spins connected to the impurity with the horizontal axis (angle $\alpha$) as obtained from the
Monte Carlo (squares) and compared with a simple variational criteria where only the spins connected to the impurity are tilted (line).
Inset: distortion angle $\alpha$ as a function of the distance from the impurity (in units of the lattice spacing $a_0$) obtained
by Monte Carlo for $J_2/J_1=0.55$.}
\label{fig1}
\end{center}
\end{figure}
We first discuss the symmetry of the magnetic order for the case of a single impurity.
The ground state for $J_2/J_1>0.5$ in the absence of impurity is continuously degenerate and is characterised
by a bi-partite lattice, with two distinct anti-ferromagnetically ordered
states on respectively each sub-lattice, and $\theta$ is the angle between the two magnetic directions.
Our Monte Carlo calculations show that introducing a single
impurity lifts the former continuous degeneracy, and selects the state with $\theta=90^\circ$, as shown in Fig.~\ref{fig1}.\textbf{a}, and in agreement with a prediction by Chris Henley\cite{henley_j1j2_dilution},
who suggested the name {\it anticollinear} to describe the state with $\theta=90^\circ$.
This state was also suggested as a stable phase of ferro-pnictides \cite{pnictides_ilya_frustration_is_large}.

This selection is induced by a local energy optimization around the impurity site.
Indeed, the spins on the impurity's nearest neighbor sites
slightly distort in order to align ferromagnetically, in order to optimize locally the energy with their own neighbors once an impurity is introduced into the $90^\circ$ state.
We show in Fig.~\ref{fig1}.\textbf{b} the obtained deviation of the spins connected to the impurity from the bulk $90^\circ$ magnetic state.
We obtain, as shown in the inset of Fig.~\ref{fig1}.\textbf{b}, that the bonds affected by this local distortion are in the very near vicinity of the
impurity. Noteworthy, treating the distortion angle $\alpha$ as a simple variational parameter, and neglecting the distortion of the spins
which are not nearest neighbors of the impurity, lead to a very good estimate of the Monte Carlo result.
The good agreement between the two methods confirms that the energy optimization around a single impurity is essentially local in space,
and thus the selection of the $90^\circ$ magnetic state is driven by a local energetic optimization process,
in contrast to the \emphasize{order by disorder} selection of the ${\bf Q}=(0,\pi),(\pi,0)$ states, which is an entropic selection.

%%%%%%%%%%%%%%%%%%%%%%%%%%%%%%%%%%%%%%%%%%%%%%%%%%%%%%%%%%%%%%%%
%%%%%%%%%%%%%%%%%%%%%%%%%%%%%%%%%%%%%%%%%%%%%%%%%%%%%%%%%%%%%%%%
%%%%%%%%%%%%%%%%%%%%%%%%%%%%%%%%%%%%%%%%%%%%%%%%%%%%%%%%%%%%%%%%
%%%%%%%%%%%%%%%%%%%%%%%%%%%%%%%%%%%%%%%%%%%%%%%%%%%%%%%%%%%%%%%%
%%%%%%%%%%%%%%%%%%%%%%%%%%%%%%%%%%%%%%%%%%%%%%%%%%%%%%%%%%%%%%%%
%%%%%%%%%%%%%%%%%%%%%%%%%%%%%%%%%%%%%%%%%%%%%%%%%%%%%%%%%%%%%%%%
%%%%%%%%%%%%%%%%%%%%%%%%%%%%%%%%%%%%%%%%%%%%%%%%%%%%%%%%%%%%%%%%
%%%%%%%%%%%%%%%%%%%%%%%%%%%%%%%%%%%%%%%%%%%%%%%%%%%%%%%%%%%%%%%%
%%%%%%%%%%%%%%%%%%%%%%%%%%%%%%%%%%%%%%%%%%%%%%%%%%%%%%%%%%%%%%%%
%%%%%%%%%%%%%%%%%%%%%%%%%%%%%%%%%%%%%%%%%%%%%%%%%%%%%%%%%%%%%%%%
%%%%%%%%%%%%%%%%%%%%%%%%%%%%%%%%%%%%%%%%%%%%%%%%%%%%%%%%%%%%%%%%
%%%%%%%%%%%%%%%%%%%%%%%%%%%%%%%%%%%%%%%%%%%%%%%%%%%%%%%%%%%%%%%%

Next, we investigate the relevance of this state as a function
of temperature and impurity concentration.
This anticollinear order is characterized similarly: $M_{90}(x)=
\norm{(\hat{{\bf {S}}}_{i} - \hat{{\bf {S}}}_{k}) \wedge (\hat{{\bf {S}}}_{j} - \hat{{\bf {S}}}_{l})}$.
Interestingly, the deviation is large for $J_2 \approx J_1/2$, and decreases significantly
as $J_2$ increases. The latter is easily explained since, when $J_2$ become large, the four nearest neighbour spin of the impurity
get antiferromagnatically aligned, and hence the distortion $\alpha$ is not energetically favored.
Interestingly, the energy optimization $\Delta E = E(\alpha)-E(\alpha=0) \approx -0.7 J_1$ is quite significant around $J_2/J_1=0.5$ and hence we expect the $90^\circ$ magnetic phase to be a relevant
magnetic phase for pnictides, since first-principle calculations obtain values for $J_2/J_1$ close to $0.5$ (such as  $J_2/J_1 \sim 0.75$ for \se \cite{spin_disorder_electronic_structure}).
\begin{figure}
\begin{center}
\includegraphics[width=1\columnwidth]{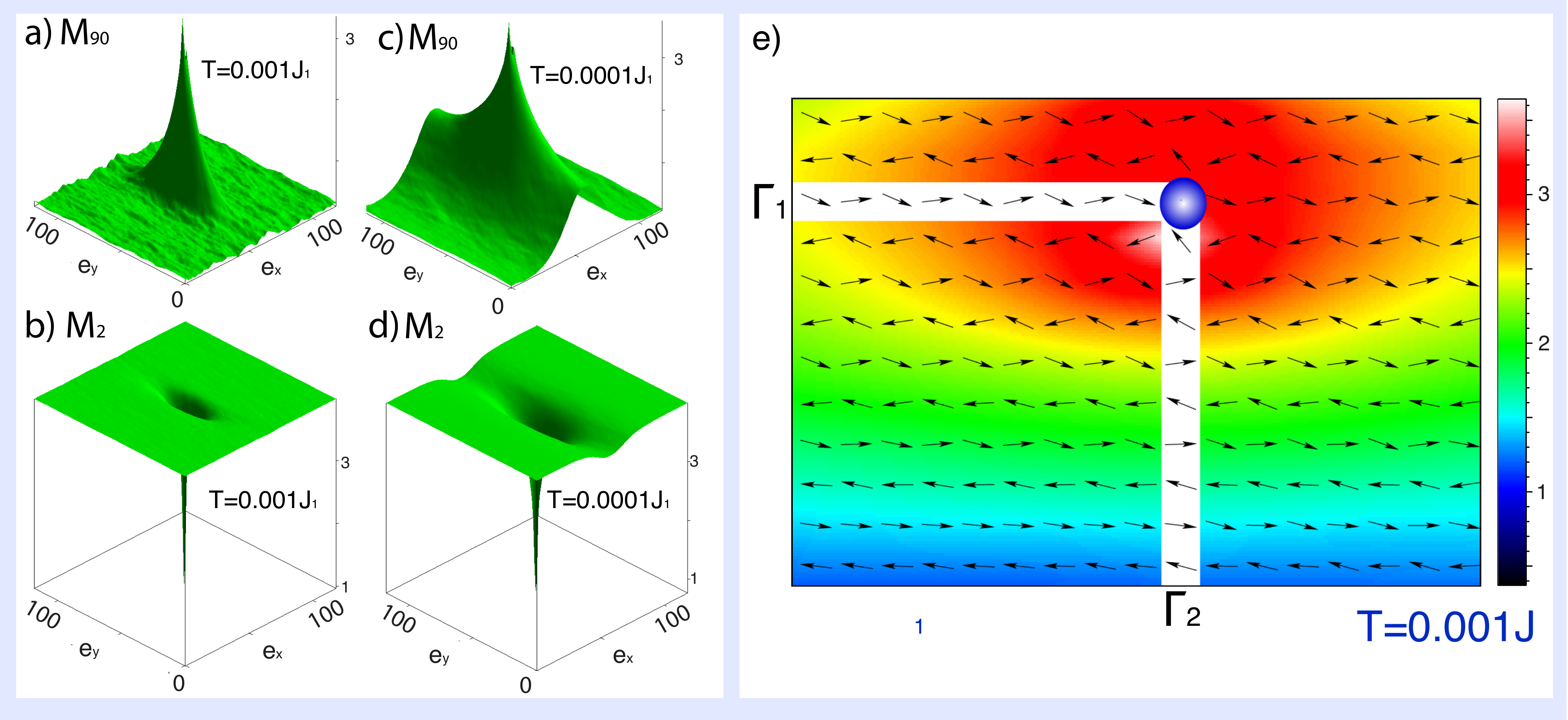}
\caption{
(Color online) Spatially resolved order parameter a) for the $90^\circ$ spin order ($M_{90}(x)$) and b) for the Ising order ($M_2(x)$) obtained at $T=10^{-3}J_1$.
The respective order parameters evaluated at a lower temperature $T=10^{-4}J_1$ are shown in c) and d). In all the calculations above the cluster contain
$L \times L=120 \times 120$ sites and the impurity is located at the center of the cluster at $(x,y)=(60,60)$. e) Color plot of $M_{90}(x)$ at  $T=10^{-3}J_1$ (blue and red
are respectively the minimum and maximum). All calculations above are done for $J_2/J_1=0.55$.
}
\label{fig2}
\end{center}
\end{figure}
We now turn to the discussion of the single impurity problem at small but finite temperature.
The $90^\circ$ spin order around a single impurity is shown in Fig.~\ref{fig2}.\textbf{a,c}.
Indeed, the $90^\circ$ order is not breaking a discrete symmetry, such as the Ising symmetry broken by the Z$_2$ Ising parameter
defined in equation (\ref{isingvar}), and is not a stable thermodynamic phase.
Thus, we observe that the $90^\circ$ order does not develop long-range correlations but is rather stabilized around the impurity within a finite region,
which defines its correlation distance. At finite temperature, there is a competition in the
free energy $F=E-TS$ between, on the one hand, the local energy optimization in the vicinity of the impurity which favors the $90^\circ$ spin order, and on the other hand
the entropy selection which favors the Ising ${\bf Q}=(0,\pi),(\pi,0)$ states. Notwithstanding that the $90^\circ$ state is energetically favored and is stabilized at short distances from the impurity, we find that thermal fluctuations screen the impurity at long distances such that the system recovers the entropically selected Ising states far from the impurity.
Hence, there are mainly three different spatial scales in this problem: i) The direct proximity of the impurity, where the $90^\circ$ state distorts to optimize the energy ($\xi_1$), ii) the correlation length of the $90^\circ$ state for which this order survives the entropic selection ($\xi_2$), iii) and finally the correlation length of the Ising state ($\xi_3$).
Furthermore, the interplay between the spatial scales is temperature dependent, in particular the
size of the $90^\circ$ cluster, as seen from Fig.~\ref{fig2}.\textbf{a} obtained at $T=10^{-3}J_1$ and  Fig.~\ref{fig2}.\textbf{c} obtained
at a smaller temperature $T=10^{-4}J_1$. In addition, we note that the shape of the $90^\circ$ cluster is highly asymmetric.
To understand this point, we measured the Ising order parameter $M_2$ in the same calculation (see Fig.~\ref{fig2}.\textbf{b,d}),
and as expected we observe a concomitant reduction of the Ising order in the region where the $90^\circ$ order is large.
Interestingly, we find that the shape of the $90^\circ$ cluster correlates with the Ising order: the latter is an
ellipsoid (the large axis is denoted by $\Gamma_1$ and the small axis by $\Gamma_2$), we find that $\Gamma_1$ is along the $\bold{e}_y$ ($\bold{e}_x$)
direction for the corresponding $Z_2=+1$ ($Z_2=-1$) Ising state. In particular, as shown in Fig.~\ref{fig2}.\textbf{e}, the spins along $\Gamma_1$ ($\Gamma_2$), highlighted
by the white stripes, corresponds to the ferromagnetically (anti-ferromagnetically ) aligned spins of the colinear Ising states. Hence, due to the antiferromagnetic $J_1$ coupling,
the deviations from the pure Ising state are energetically favorable along $\Gamma_1$ , and costly along $\Gamma_2$, which explains why the screening
of the impurity is more effective in one direction than the other.

%%%%%%%%%%%%%%%%%%%%%%%%%%%%%%%%%%%%%%%%%%%%%%%%%%%%%%%%%%%%%%%%
%%%%%%%%%%%%%%%%%%%%%%%%%%%%%%%%%%%%%%%%%%%%%%%%%%%%%%%%%%%%%%%%
%%%%%%%%%%%%%%%%%%%%%%%%%%%%%%%%%%%%%%%%%%%%%%%%%%%%%%%%%%%%%%%%
%%%%%%%%%%%%%%%%%%%%%%%%%%%%%%%%%%%%%%%%%%%%%%%%%%%%%%%%%%%%%%%%
%%%%%%%%%%%%%%%%%%%%%%%%%%%%%%%%%%%%%%%%%%%%%%%%%%%%%%%%%%%%%%%%
%%%%%%%%%%%%%%%%%%%%%%%%%%%%%%%%%%%%%%%%%%%%%%%%%%%%%%%%%%%%%%%%

\begin{figure}
\begin{center}
\includegraphics[width=1.0\columnwidth]{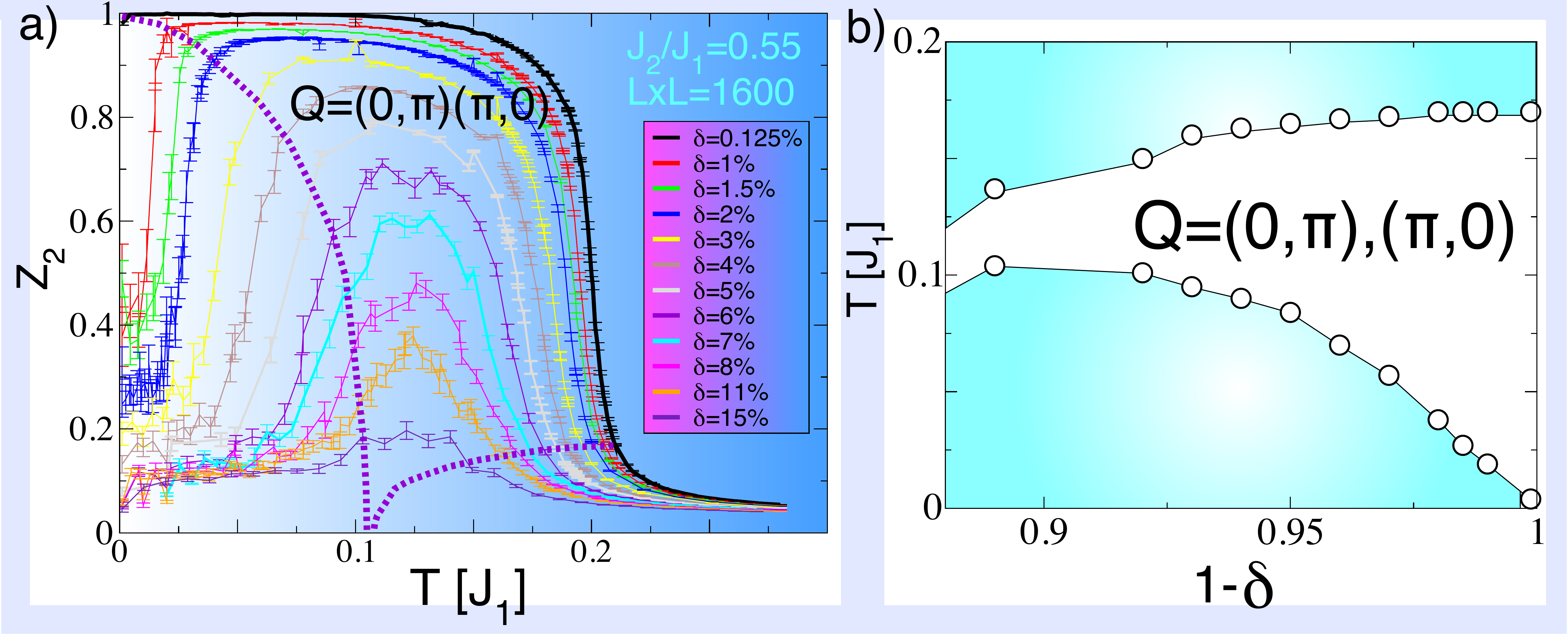}
\caption{
(Color online) a) Temperature dependence of the spatially averaged Ising order parameter $Z_2$ for various
impurity concentrations $\delta$ obtained for a $L \times L = 40 \times 40$ lattice.
The Ising order is suppressed at low temperature by the presence of impurities, and at large temperature
by strong thermal fluctuations. The dashed lines are guide to the eyes to track the Ising crossovers at low
and high temperatures. b) Corresponding phase diagram in impurity density $\delta$ and temperature T.
All calculations above are done for $J_2/J_1=0.55$ and were averaged over 32 random configurations of disorder.}
\label{fig3}
\end{center}
\end{figure}
Let us now extend the discussion to a finite concentration of impurities $\delta$ (see Fig.~\ref{fig3}.\textbf{a}).
We carried out Monte Carlo calculations for a $L \times L = 40 \times 40$ cluster with different
magnetic impurities dilutions $\delta$, and we averaged the physical observables over 32 random configurations of impurities.
We find for small concentration $\delta=0.125\%$ only a weak
effect on the Ising order. In particular, we observe the Ising-like cross-over at $T \approx 0.2 J_1$,
as shown by the sharp drop of the order parameter as this temperature,
and a very small dip in the $Z_2$ order parameter at $T \approx 0.005 J_1$.
We note that transitions belonging to the Ising universality class ($\nu=1$ and the dimension $d=2$) do not satisfy the well known \emphasize{Harris criteria}
\cite{harris_criteria}, which assess that phase transition with $\nu d >2$ are unaffected by the disorder.
However, as numerical studies show that the 2D Ising model is weakly affected by the disorder~\cite{disorder_2d_ising}, our results show
that the cross-over temperature at $T \approx 0.2 J_1$ is not strongly affected by moderate disorder, and hence suggest that the phase transition
survives in the presence of impurities. A formal proof would require a detailed finite size scaling analysis and goes beyond the scope
of this work.

At larger concentrations $0.125\%<\delta<11\%$, we find a steady decrease of the Ising order at small temperatures,
as highlighted by the dashed lines in Fig.~\ref{fig3}.\textbf{a}. For instance, at $\delta=2\%$ we observe a quench of the Ising
order for $T < 0.025 J_1$, and the entropic selection kicks in for $0.025 < T < 0.19$ where we obtain the Ising ordered phase,
and finally at higher temperatures  $t>0.19$ the system is a paramagnet. This scenario is commonly called \emphasize{re-entrance phase transition}.
We note that the impurities mainly affect the Ising order at low temperatures, and the Ising-like transition near $T \approx 0.2 J_1$ is moderately affected
by impurities at small and moderate dilutions. Beyond a critical dilution $\delta_c \approx 20\%$, we do not observe the presence of the collinear or anticollinear states.
Indeed, in two dimensions and for an impurity dilution $\delta_c=1/9$, there is one impurity in average connected to every spin of the lattice, so that the
local distortions and subsequent local energy optimizations start to prevail over both the Ising phase and the $90^\circ$
local spin order.
The phase diagram is summarized in Fig.~\ref{fig3}.\textbf{b}.
Note that the reentrant behavior of the Ising phase agrees with the prediction of Ref.~\onlinecite{henley_j1j2_dilution}. However, the rest of the phase diagram of Ref.~\onlinecite{henley_j1j2_dilution} cannot be compared to the present results. Indeed, in the case of the XY model studied in Ref.~\onlinecite{henley_j1j2_dilution}, the local chirality of the anticollinear order defines an
Ising variable, and the anticollinear phase must be separated from the paramagnetic phase by a phase
transition. By contrast, no long-range order can exist for the vector chirality at finite temperature
for Heisenberg spins, and the low-temperature phase below the Ising phase can be smoothly connected to the
paramagnetic phase. To study this crossover would require to study larger dilutions and to perform
a systematic finite size scalings, which goes beyond the scope of the present work.

%%%%%%%%%%%%%%%%%%%%%%%%%%%%%%%%%%%%%%%%%%%%%%%%%%%%%%%%%%%%%%%%
%%%%%%%%%%%%%%%%%%%%%%%%%%%%%%%%%%%%%%%%%%%%%%%%%%%%%%%%%%%%%%%%
%%%%%%%%%%%%%%%%%%%%%%%%%%%%%%%%%%%%%%%%%%%%%%%%%%%%%%%%%%%%%%%%
%%%%%%%%%%%%%%%%%%%%%%%%%%%%%%%%%%%%%%%%%%%%%%%%%%%%%%%%%%%%%%%%
%%%%%%%%%%%%%%%%%%%%%%%%%%%%%%%%%%%%%%%%%%%%%%%%%%%%%%%%%%%%%%%%
%%%%%%%%%%%%%%%%%%%%%%%%%%%%%%%%%%%%%%%%%%%%%%%%%%%%%%%%%%%%%%%%

\begin{figure}
\begin{center}
\includegraphics[width=1.0\columnwidth]{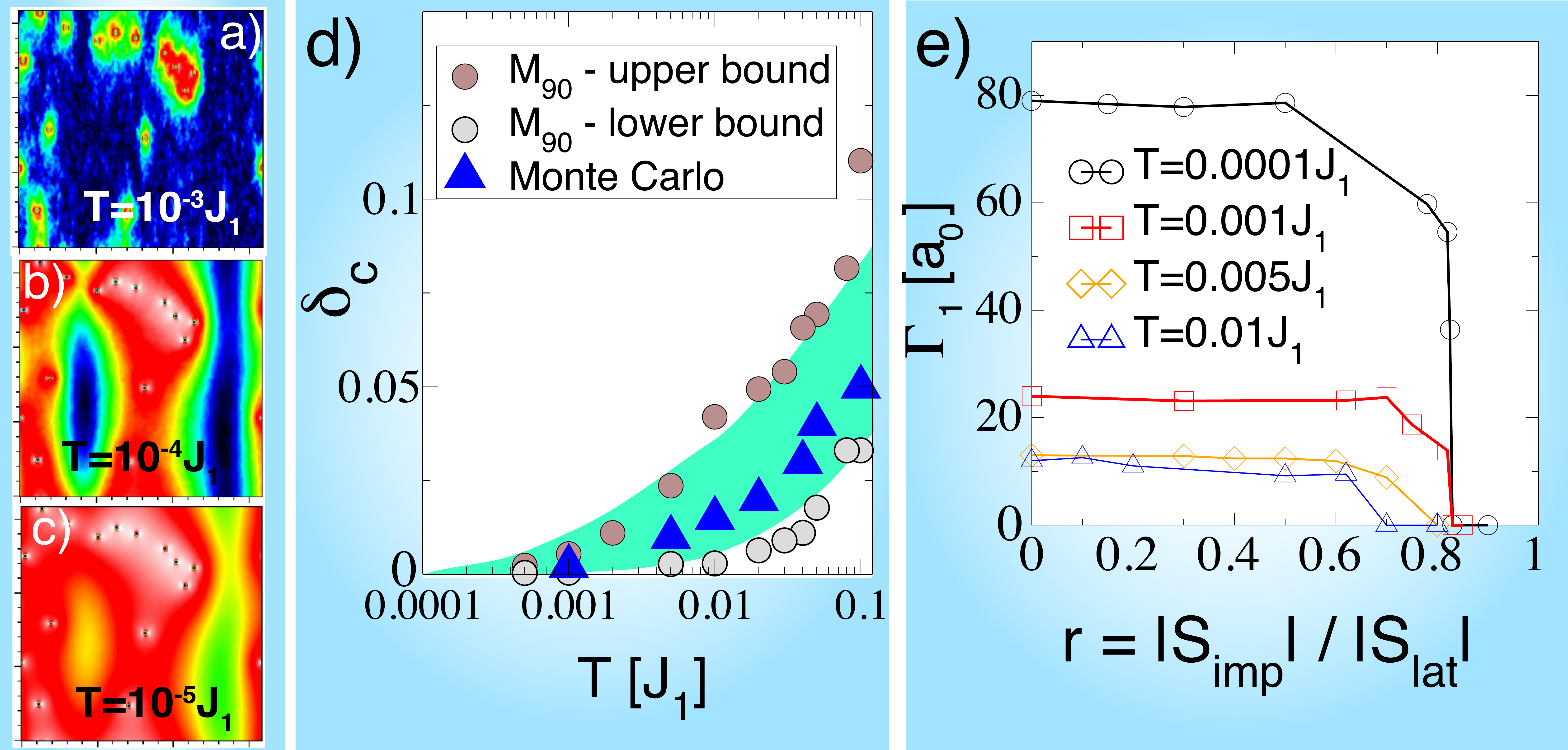}
\caption{
(Color online) Real space map of $M_{90}$ obtained at a) $T= 10^{-3} J_1$, b) $T= 10^{-4} J_1$ and c) $T= 10^{-5} J_1$ for
a given configuration of $16$ non-magnetic impurities in a $L \times L = 120 \times 120$ lattice ($\delta=0.1\%$).
The colours range from black (minimum) to red (maximum). d) Critical dilution
$\delta_c$ as a function of the temperature obtained by the Monte Carlo data of Fig.~\ref{fig3}.\textbf{a} (triangles), and by a simple criteria
comparing the length of the short (dark gray circles) and long (light gray circles) ellipsoidal axis of the $90^\circ$ order cluster (see discussion in the text).
e) $\Gamma_1$ (the long axis of the 90$^\circ$ order ellipsoid) in units of $a_0$ as a function of $r=S_{imp}/S_{lat}$, where $S_{imp}$ is
the spin of the magnetic impurity, and $S_{lat}$ is the spin of the correlated element of the compound. All calculations above are done for $J_2/J_1=0.55$.}
\label{fig4}
\end{center}
\end{figure}
Interestingly enough, the re-entrance phenomena observed in Fig.~\ref{fig3}.\textbf{a} can be explained
at small impurity concentration on the basis of the single impurity results at finite temperature.
Let us consider a given configuration of impurities for a finite impurity dilution
of $\delta=0.01$ at  $T= 10^{-3},10^{-4},10^{-5} J_1$ (see Fig.~\ref{fig4}.\textbf{a,b,c}). The $90^\circ$ order stays
localized around the impurities at high temperature (Fig.~\ref{fig4}.\textbf{a}), but forms superstructures connecting the impurities (Fig.~\ref{fig4}.\textbf{b}) upon lowering the temperature
until it finally spreads through the whole lattice (Fig.~\ref{fig4}.\textbf{c}). This process is very similar to a percolation transition, and can be captured within a
very simple argument: The $90^\circ$ order become long-range when the $90^\circ$ order correlation length $\xi_{90}(T)$ is of the order of the mean distance $\lambda$
between the impurities. We note that the  $90^\circ$ is an ellipsoid, and hence we obtain a lower and upper bound on the critical dilution, by comparing the short and
long axis respectively to the mean impurity-impurity distance.
Indeed, we compare in Fig.~\ref{fig4}.\textbf{d} the critical dilution obtained from the Monte Carlo data of Fig.~\ref{fig3}.\textbf{a} and with the critical dilution obtained by this
simple argument. We find that this argument provides a reliable estimate of the critical dilution. This confirms that the re-entrance of the Ising order is
driven by the competition of different length scales associated to the Ising and $90^\circ$ order parameters.

Finally, we generalized our calculations to magnetic impurities with non zero spin (see Fig.~\ref{fig4}.\textbf{e}). Remarkably, we find that the
$90^\circ$ order cluster around a single impurity is not significantly affected by the spin of the impurity $S_{imp}$, as long as the ratio of the spin of the
impurity to the magnetic element of the compound $r=S_{imp}/S_{lat}$ remains smaller than $\approx 0.6$.
This suggests that the re-entrance phase transition does not strictly require non-magnetic impurities but
could also be present for instance in the case of Ni impurities in \ba,
where the spin of Ni is $\approx 40\%$ of the spin of Fe\cite{j1j2_disorder_our_paper_n_curro}. So it will be very
interesting to see if the order proposed in this paper can lead to an alternative interpretation of the
NMR results in \ba and maybe help clarifying the origin of the lineshapes in Ni and Zn doped samples \cite{j1j2_disorder_our_paper_n_curro}.

%%%%%%%%%%%%%%%%%%%%%%%%%%%%%%%%%%%%%%%%%%%%%%%%%%%%%%%%%%%%%%%%
%%%%%%%%%%%%%%%%%%%%%%%%%%%%%%%%%%%%%%%%%%%%%%%%%%%%%%%%%%%%%%%%
%%%%%%%%%%%%%%%%%%%%%%%%%%%%%%%%%%%%%%%%%%%%%%%%%%%%%%%%%%%%%%%%
%%%%%%%%%%%%%%%%%%%%%%%%%%%%%%%%%%%%%%%%%%%%%%%%%%%%%%%%%%%%%%%%
%%%%%%%%%%%%%%%%%%%%%%%%%%%%%%%%%%%%%%%%%%%%%%%%%%%%%%%%%%%%%%%%
%%%%%%%%%%%%%%%%%%%%%%%%%%%%%%%%%%%%%%%%%%%%%%%%%%%%%%%%%%%%%%%%

In conclusion, we have carried out a systematic study of the effect of non-magnetic impurities in a frustrated Heisenberg model.
We reported that for $J_2/J_1>0.5$ the continuous degeneracy of the ground state induced by the frustration
is lifted due to a local optimization in the vicinity of a single impurity. The energy gain favors anticollinear order,
which consists of bipartite lattices supporting N\'eel states entangled with a $90^\circ$ angle. This order, energetically favored,
competes with the Ising order, entropically favored, and at long distance we find that the impurity is screened by thermal fluctuations. This results in a rich phase diagram with the stabilization of anticollinear order at low temperature as soon as a finite concentration of impurities is present followed by a reentrant collinear phase upon increasing the
temperature. Moreover, we have shown that the structure around the impurity locally departs from the purely anticollinear order. This effect is large when $J_2/J_1$ is close to $1/2$, as in the pnictides, and should
be detectable by local probes such as NMR.

We would like to thank N. Curro for very useful explanations about the NMR results of Ref.~\onlinecite{j1j2_disorder_our_paper_n_curro}, and F. Becca and A. L\"auchli for
interesting discussions at an early stage of the project. We are especially indebted to I. Eremin 
and A. L\"auchli for quite insightful suggestions following their critical reading of the manuscript. 
C.W. was supported by the Swiss National Foundation for Science (SNFS).
F.M. is supported by the Swiss National Fund and by MaNEP.

%%%%%%%%%%%%%%%%%%%%%%%%%%%%%%%%%%%%%%%%%%%%%%%%%%%%%%%%%%%%%%%%
%%%%%%%%%%%%%%%%%%%%%%%%%%%%%%%%%%%%%%%%%%%%%%%%%%%%%%%%%%%%%%%%
%%%%%%%%%%%%%%%%%%%%%%%%%%%%%%%%%%%%%%%%%%%%%%%%%%%%%%%%%%%%%%%%
%%%%%%%%%%%%%%%%%%%%%%%%%%%%%%%%%%%%%%%%%%%%%%%%%%%%%%%%%%%%%%%%
%%%%%%%%%%%%%%%%%%%%%%%%%%%%%%%%%%%%%%%%%%%%%%%%%%%%%%%%%%%%%%%%
%%%%%%%%%%%%%%%%%%%%%%%%%%%%%%%%%%%%%%%%%%%%%%%%%%%%%%%%%%%%%%%%
%%%%%%%%%%%%%%%%%%%%%%%%%%%%%%%%%%%%%%%%%%%%%%%%%%%%%%%%%%%%%%%%
%%%%%%%%%%%%%%%%%%%%%%%%%%%%%%%%%%%%%%%%%%%%%%%%%%%%%%%%%%%%%%%%
%%%%%%%%%%%%%%%%%%%%%%%%%%%%%%%%%%%%%%%%%%%%%%%%%%%%%%%%%%%%%%%%
%%%%%%%%%%%%%%%%%%%%%%%%%%%%%%%%%%%%%%%%%%%%%%%%%%%%%%%%%%%%%%%%
%%%%%%%%%%%%%%%%%%%%%%%%%%%%%%%%%%%%%%%%%%%%%%%%%%%%%%%%%%%%%%%%
%%%%%%%%%%%%%%%%%%%%%%%%%%%%%%%%%%%%%%%%%%%%%%%%%%%%%%%%%%%%%%%%
%%%%%%%%%%%%%%%%%%%%%%%%%%%%%%%%%%%%%%%%%%%%%%%%%%%%%%%%%%%%%%%%
%%%%%%%%%%%%%%%%%%%%%%%%%%%%%%%%%%%%%%%%%%%%%%%%%%%%%%%%%%%%%%%%
%%%%%%%%%%%%%%%%%%%%%%%%%%%%%%%%%%%%%%%%%%%%%%%%%%%%%%%%%%%%%%%%
%%%%%%%%%%%%%%%%%%%%%%%%%%%%%%%%%%%%%%%%%%%%%%%%%%%%%%%%%%%%%%%%

\bibliographystyle{apsrev4-1}
%\bibliography{./biblio_cedric}
\bibliography{/Users/cweber/Documents_Cedric/BIBLIO/biblio_cedric}
\end{document}